\documentclass{iau}

\usepackage{amsmath}
\usepackage{graphicx}
\usepackage{multirow}
\usepackage{comment}
\usepackage{subcaption}

\begin{document}

\lefttitle{Devojyoti Kansabanik}
\righttitle{Heliopolarimetry}

\jnlPage{1}{7}
\jnlDoiYr{2024}
\doival{10.1017/xxxxx}
\volno{388}
\pubYr{2024}
\journaltitle{Solar and Stellar Coronal Mass Ejections}

\aopheadtitle{Proceedings of the IAU Symposium}
\editors{N. Gopalswamy,  O. Malandraki, A. Vidotto \&  W. Manchester, eds.}

\title{Preparing for Heliopolarimetry using New-generation Ground-based Radio Telescopes}

\author{Devojyoti Kansbanik$^{1}\footnote{NASA Jack Eddy fellow hosted at the Johns Hopkins University Applied Physics Laboratory}$, Angelos Vourlidas$^2$}
\affiliation{$^1$Cooperative Programs for the Advancement of Earth System Science, University Corporation for Atmospheric Research, Boulder, CO, USA}
\affiliation{$^2$The Johns Hopkins University Applied Physics Laboratory, 11101 Johns Hopkins Road, Laurel, MD 20723, USA}

\begin{abstract}
Coronal mass ejections (CMEs) are large-scale ejections of magnetized plasma from the Sun and are associated with the most extreme space weather events. The geoeffectiveness of a CME is primarily determined by the southward component of its magnetic fields (CME-$B_z$). Recent studies have shown that CMEs evolve significantly in the inner heliosphere ($\sim20-90\ R_\odot$), and relying on extrapolations from low coronal heights can lead to wrong predictions of CME-$B_z$ in the vicinity of Earth. Hence, it is important to measure CME magnetic fields at these heights to improve CME-$B_z$ prediction. A promising method to measure the CME-entrained magnetic field in the inner heliosphere is by measuring the changes in Faraday rotation (FR) of linearly polarized emission from background radio sources as their line-of-sight crosses the CME plasma. Here, we present the current preparation of new-generation ground-based radio telescopes for this purpose. 
\end{abstract}


\maketitle

\section{Introduction}
Coronal mass ejections (CMEs) are large-scale expulsions of magnetized plasma from the Sun and the most important driver for space weather (SpWx). The geoeffectiveness of a CME is primarily determined by the entrained magnetic field strength and topology, particularly the southward component (CME-$B_z$), the direction relative to the dipolar magnetic field of the Earth. Long periods of southward $B_z$ lead to the opening of the magnetosphere and the injection of energy and solar plasma into the terrestrial system which in turn disturbs the system in multiple ways, collectively known as space weather. CMEs are routinely observed by white-light coronagraphs and heliospheric imagers at visible wavelengths, which capture the density structure of the CMEs. These decades-long observations have provided several geometrical and dynamical properties of the CMEs building our current understanding of CMEs \citep{Howard2023}. However, the white light observations cannot provide a direct measure of the CME-entrained vector magnetic fields.

Observing techniques using modern radio telescopes provide some means to estimate the magnetic field of CMEs at coronal heights based on different types of radio emission (e.g. gyrosynchrotron and gyroresonance emission) \citep{Carley2020,Vourlidas2020}. However, beyond about $\sim$10 $R_\odot$ ($R_\odot$= solar radii), the optimal observing frequency for these emissions falls below the ionospheric cutoff ($<$10 MHz). Since the strength of these emissions is generally too low for detection by space-based non-imaging radio instruments, this approach is only useful in the lower corona. However, interactions between different CMEs, ambient solar wind, and other heliospheric structures (like corotating or stream interaction regions) can lead to structural changes of CMEs in the inner heliosphere ($\sim$20-90 $R_\odot$) \citep{Manchester2017,Braga_2022}. These deformations complicate the prediction of CME arrival times and CME-$B_z$ at 1 AU from the information obtained when the CME was at lower coronal heights \citep{Vourlidas2019}. Hence, the measurement of CME magnetic fields at higher coronal heights and inner heliosphere is essential for improving SpWx forecasting and predicting the geo-effectiveness of CMEs.

Measurement of changes in Faraday rotation \citep[FR,][]{Ferri2021} of background linearly polarized radio sources due to coronal and heliospheric plasma \citep[see][for a review]{Kooi2022} is known to be a promising remote-sensing tool for measuring the line-of-sight (LoS) magnetic field both in the corona and the inner heliosphere. We have coined the term, ``Heliopolarimetry" to describe this observation technique. However, its application to SpWx research remains limited due to observational and modeling challenges. 

In this proceeding, we first review the current state of heliopolarimetry in Section \ref{sec:current_status}. This will be followed by the capabilities of new-generation ground-based radio telescopes to perform heliopolarimetry at different solar elongations (Section \ref{sec:new_instrument}) and the challenges involved (Section \ref{sec:challenges}). We discuss the ongoing work for preparing these new-generation instruments for heliopolarimetry in Section \ref{sec:dis_future}. 

\section{Heliopolarimetry: Current Status}\label{sec:current_status}
FR is the rotation of the plane of polarization of linearly polarized emission as it passes through magnetized plasma. FR is given by $\lambda^2\times$ RM, where the rotation measure (RM) \citep{Brentjens2005} is the LoS integration of the product of electron density and LoS magnetic fields (RM = $\int n_e \Vec{B}. d\Vec{s}$) and $\lambda$ is the wavelength of the radiation. White-light observations of Thomson scattered photospheric emission from the electrons in the CME plasma \citep{Inhester2015}, observed using coronagraphs and heliospheric imagers, can provide the electron column density of a LoS, $N_e = \int n_e ds$. Combining both of these measurements can provide an estimate of the LoS averaged magnetic field ($<B>=\frac{RM}{N_e}$) of the CME. 

Most of the previous FR experiments either used satellite beacons during short and infrequent periods of favorable observing geometry \citep[e.g.,][]{Bird1985,Jensen2013,Wexler2019} or used high-frequency observations of a set of galactic/extragalactic linearly polarized radio sources \citep{Kooi2017,Kooi2021} mostly observing with the Jansky Very Large Array \citep[JVLA,][]{Kellermann2020}. JVLA was primarily used due to its well-characterized polarization performance. These experiments were limited to coronal heights ($<15\ R_\odot$) and could only provide sparse and temporally infrequent FR measurements due to the small field of view (FoV) of the instrument. Hence, the inversion of vector magnetic fields from the non-co-temporal measurements of $<B>$ invariably requires several assumptions and simplified models of the CME, which limits the benefit of heliopolarimetry for SpWx research and application.

To extend the applicability of heliopolarimetry for SpWx research, we need to expand the observation at higher coronal heights and into the inner heliosphere. As the CME propagates out in the heliosphere, both its electron density and magnetic fields decrease as well as its RM value. For a typical CME, the RM contribution by the CME drops from 10\,rad/m$^{2}$ at $<2^\circ$ to 0.1\,rad/m$^{2}$ at 90 $R_\odot$ \citep{Oberoi2012}. Since FR is proportional $\lambda^2$, at low frequency, small changes in RM produce detectable changes in FR. Hence, to achieve the required RM precision for heliopolarimetry at higher corona and inner heliosphere, one needs low-frequency ($\sim80-1000$ MHz) observations. CME also expands in size as it propagates and covers a larger sky area. Wide FoV instruments are required to perform heliopolarimetry on a significant part of the CME. 

\section{Capabilities of New Generation Ground-based Radio Telescopes}\label{sec:new_instrument}
Over the last decade, several new-generation radio interferometric radio telescopes have become operational, which provide some basic requirements of heliopolarimetry, wide FoV, wide frequency coverage from low to high frequency, polarization purity of the instrument, and good sensitivity over small temporal integration. These telescopes include -- the Murchison Widefield Array \citep[MWA,][]{Tingay2013,Wayth2018}, LOw Frequency ARray \citep[LOFAR,][]{lofar2013}, MeerKAT \citep{meerkat2016} and Australian Square Kilometre Array Pathfinder \citep[ASKAP,][]{Hotan2021}. 
\begin{figure*}
    \centering
    \includegraphics[scale=0.5]{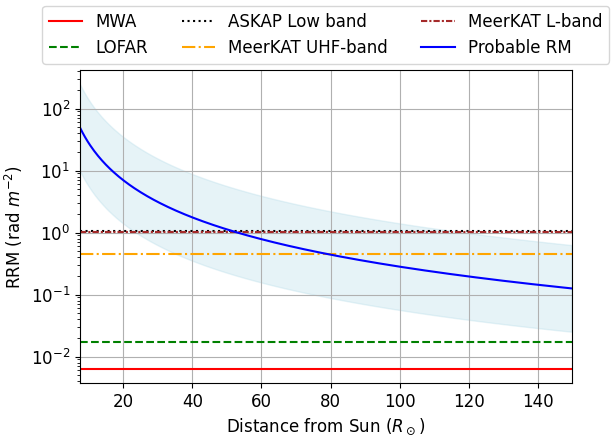}
    \caption{Probable variation of RRM of a CME with heliocentric distances is shown by the solid blue line, with an order of magnitude variance by light blue shade. Different horizontal lines represent the achievable precision in RRM with $\sim15$ minutes of integration for different instruments in different observing bands.}
    \label{fig:probable_RM}
    \vspace{-0.3cm}
\end{figure*}

Each of these instruments operates at different frequency ranges with different observing bandwidths. To measure the RM contribution from the CME, we need precision in relative rotation measure (RRM = RM$_\mathrm{obs}$ - RM$_\mathrm{source}$) more than the RM of the CME. For a given instrument and frequency range, the maximum precision in the RRM is given as $\delta\text{RRM} \sim\frac{\delta\phi_{\text{FWHM}}}{\sqrt{2}\text{SNR}}$. $\delta\phi_{\text{FWHM}} = 2\sqrt{3}/\Delta\lambda^2$ is the half maximum of the full width of the rotation measure synthesis function \citep{Brentjens2005}, where $\Delta\lambda^2$ is the wavelength squared span across the observing bandwidth.   
\begin{figure*}
    \centering
    \includegraphics[scale=0.4]{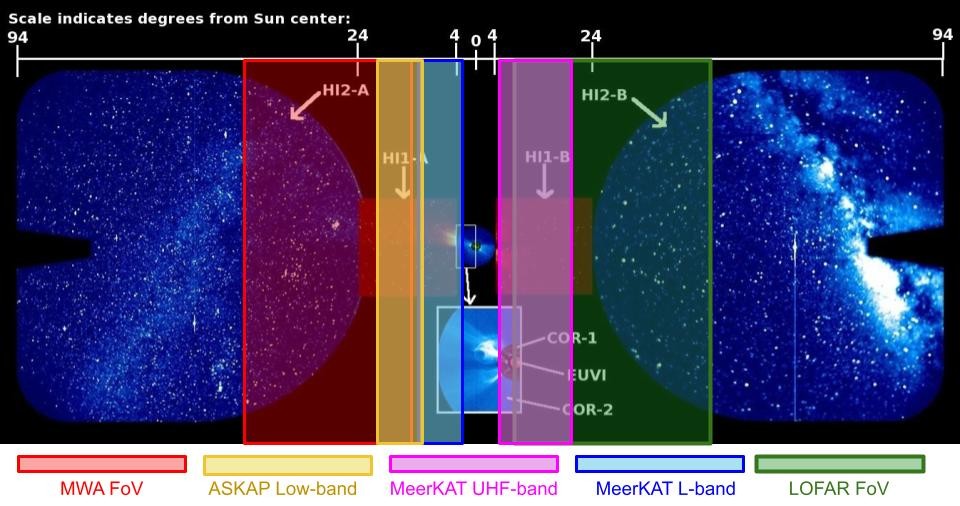}
    \caption{The ranges of heliocentric distances suitable for heliopolarimetry using the new-generation wide FoV instruments. The spans are only shown along the ecliptic plane, while the same is true for in any direction from the Sun. The background image is composed of images from coronagraph and heliospheric imagers of STEREO/SECHHI.}
    \label{fig:fovs}
    \vspace{-0.3cm}
\end{figure*}

\begin{table*}[!ht]
\centering
    \renewcommand{\arraystretch}{1.4}
    \begin{tabular}{|p{2.4cm}|p{3cm}|p{2.2cm}|p{2.2cm}|}
    \hline
       Telescope Name & Frequency coverage & Minimum $R_\odot$ & Maximum $R_\odot$ \\ \hline \hline 
        MWA & 80 -- 300 MHz & 25 & 150\\
        \hline
        LOFAR & 110 -- 240 MHz & 20 & 150 \\
        \hline
        ASKAP-Low & 740 -- 1060 MHz & 30 & 50 \\
        \hline
        MeerKAT-UHF & 580 -- 1015 MHz & 18 & 75\\
       \hline
        MeerKAT-L & 900 -- 1670 MHz & 13 & 50\\
       \hline
    \end{tabular}
    \caption{Frequency coverage, minimum, and maximum solar elongation for each of the instruments.}
    \label{table:table1}
    \vspace{-0.5cm}
\end{table*}

In Figure \ref{fig:probable_RM}, the solid blue line shows the variation of probable RRM \citep{Jensen2010,Oberoi2012} value of a CME with the distance from the Sun. Since, at higher corona and inner heliosphere, there is no measurement available, we have shown the region of an order of magnitude variation about the mean probable RRM value by the blue shade. Different horizontal lines represent the precision in RRM that can be achieved with $\sim15$ minutes of integration of these instruments. We have taken the observing bandwidth mentioned in the second column of Table \ref{table:table1} for this calculation. We emphasize that these RRM estimates are only order-of-magnitude estimates and may be significantly different in a real scenario. The coronal and heliospheric region for which the value of probable RRM is higher than the achievable RRM precision of the instrument limits the maximum heliocentric distance of that instrument for heliopolarimetry. These values are mentioned in the fourth column of Table \ref{table:table1}. All of these instruments have demonstrated their polarimetric capability (e.g., MWA \citep{Lenc2017,Riseley2020}, LOFAR \citep{Jelić2015,Sullivan2023}, MeerKAT \citep{Taylor2024}, ASKAP \citep{Thomson2024}) and conducted several polarization sky surveys. This survey data provides a reference RM of the sky for measuring changes in RM due to CMEs. 

\section{Preparing for Challenges with New Generation Instruments}\label{sec:challenges}
While these new generation instruments are technically capable of providing high sensitivity and high precision polarimetry, the wide FoV poses some challenges when observing close to the Sun. To achieve similar sensitivity to the astronomical surveys, there are minimum solar elongations where these telescopes can be pointed. If they are pointed closer than this limit, the increase in system temperature of the instrument as well as radio emission from the Sun makes the calibration non-trivial and also reduces the sensitivity. The minimum solar elongations for these instruments at different observing bands are listed in the third column of Table \ref{table:table1}. Considering both minimum and maximum solar elongations, these instruments can make heliopolarimetry possible over $\sim15 - 150\ R_\odot$.  This region has not yet been explored but holds important clues for the CME kinematics \citep[e.g.,][]{Colaninno_Vourlidas_Wu_2013,Sachdeva_etal_2017}. The suitable heliospheric region for each instrument are shown in Figure \ref{fig:fovs}.

For observing as close as possible to the Sun, the main hurdle is to place the Sun within the first null of the primary beam. Given the chromaticity and wide observing bandwidth, it is impossible to exactly place the Sun at the null of the primary beam over the entire observing band. Hence, special care needs to be taken during calibration to account for the solar emission. The next hurdle for low-frequency observations targeting the inner heliosphere is the contribution of the ionospheric to the measured RM. The RM contribution by the CME at these heliocentric heights are comparable to ionospheric RM. Hence, a very precise correction of the ionospheric RM is necessary for the inner heliospheric RM measurements using the MWA and LOFAR.

\section{Discussion}\label{sec:dis_future}
Although these new-generation wide FoV ground-based radio telescopes are capable of heliopolarimetry, no observations and studies have been done so far. Given the increasing numbers of CME, as we approach the peak of Solar Cycle 25, this is the ideal time to start heliopolarimetric observations with these instruments. We have recently initiated observations using some of these instruments and also look at suitable archival observations to solve different calibration challenges and demonstrate the measurement of $<B>$ along multiple LoS piercing through the CME. Once these measurements are available, they will be used for constraining CME flux rope models to understand how these new constraints improve the CME-$B_z$ prediction at 1 AU. 

\begin{acknowledgements}
This research was supported by the NASA Living with a Star Jack Eddy Postdoctoral Fellowship
Program, administered by UCAR’s Cooperative Programs for the Advancement of Earth System
Science (CPAESS) under award \#80NSSC22M0097. D.K. thanks the organizers of IAU Symposium 388 for providing travel support to attend the symposium and present this work. D. K. thanks John Morgan for producing the term Heliopolarimetry.    
\end{acknowledgements}

\bibliographystyle{apj}
\bibliography{sample}

\end{document}